\documentclass[a4paper]{jpconf}
\usepackage{graphicx,amssymb,amsmath}
\usepackage[usenames,dvipsnames]{color}

\begin{document}

\title{Transverse and longitudinal vibrations in amorphous silicon.}
\author{Y M Beltukov$^{1,2}$, C Fusco$^3$, A Tanguy$^3$ and D A Parshin$^4$}
\address{$^1$A. F. Ioffe Physical-Technical Institute, 194021 Saint-Petersburg, Russia}
\address{$^2$Universit\'e Montpellier II, CNRS, Montpellier 34095, France}
\address{$^3$Universit\'e Lyon 1, F-69622 Villeurbanne Cedex, France}
\address{$^4$Saint Petersburg State Polytechnical University, 195251 Saint-Petersburg, Russia}

\begin{abstract}
We show that harmonic vibrations in amorphous silicon can be decomposed to transverse and longitudinal components in all frequency range even in the absence of the well defined wave vector ${\bf q}$. For this purpose we define the transverse component of the eigenvector with given $\omega$ as a component, which does not change the volumes of Voronoi cells around atoms. The longitudinal component is the remaining orthogonal component. We have found the longitudinal and transverse components of the vibrational density of states for numerical model of amorphous silicon. The vibrations are mostly transverse below 7 THz and above 15 THz. In the frequency interval in between the vibrations have a longitudinal nature. Just this sudden transformation of vibrations at 7 THz from almost transverse to almost longitudinal ones explains the prominent peak in the diffusivity of the amorphous silicon just above 7 THz.
\end{abstract}

\section{Introduction}

The propagation of vibrational excitations in disordered media is one of the important problems in condensed matter physics. Microscopic nature of these vibrational excitations is still poorly understood, despite that they are responsible for the specific heat, the thermal conductivity and the sound propagation in amorphous materials.

In 1999 Allen and Feldman proposed a simple classification of vibrations in disordered media~\cite{Allen-1999}, which was based on the numerical simulations of amorphous silicon. The low-frequency vibrations are longitudinal (LA) or transverse (TA) plane waves (acoustic phonons). The mean free path of phonons decreases rapidly with frequency due to structural or dynamical disorder. At some frequency $\omega_{\rm IR}\approx 4$ THz the mean free path of phonons becomes comparable with its wavelength. It is a so-called Ioffe-Regel criterion. However, the mobility edge $\omega_{\rm loc}$ for vibration localization is much higher then $\omega_{\rm IR}$ what is not the case for electron localization. Vibrations in amorphous materials in this relatively wide frequency range $\omega_{\rm IR}<\omega<\omega_{\rm loc}$ are delocalized but not propagated as phonons. These vibrations were called \emph{diffusons} because they spread by means of diffusion~\cite{Allen-1999, Beltukov-2013}.

However, this picture cannot clearly explain the prominent peak (more than 3 times in the value) in the diffusivity of vibrations in amorphous silicon between 7 and 13 THz \cite{Allen-1999, Feldman-1999}. Allen and Feldman noticed that the sharp rise of the diffusivity at 7 THz corresponds to the end of TA branch in crystalline Si which is responsible for the local minimum in the vibrational density of states. So the frequency 7 THz marks a point in the spectrum where vibrations change their character from somewhat TA-like, to somewhat LA-like with much larger group velocity. However the notions of transverse and longitudinal vibrations are ill defined for such big frequencies in amorphous silicon.

In this paper we will generalize the notions of transverse and longitudinal vibrations for disordered systems like amorphous silicon. As an example we study numerically the model amorphous silicon (a-Si) system consisting of $N=32768$ atoms contained in a periodic cubic box of lengths $L_x=L_y=L_z$ of approximately 87 \AA. The technical details of the preparation of the a-Si sample have already been presented in Ref.~\cite{b.fusco1}. The Si-Si interaction in the amorphous silicon studied here is well described by the Stillinger-Weber potential~\cite{b.stillinger}.

\section{Longitudinal and transverse components of displacements}

Low-frequency vibrations even in amorphous media are well-defined plane longitudinal and transverse waves. These vibrations can be described in continuous medium approximation. The displacement fields ${\bf u}({\bf r})$ for longitudinal (L) and transverse (T) waves have a form
\begin{gather}
    {\bf u}_{\textsc{l},\textsc{t}}({\bf r}) = {\bf u}_{\textsc{l},\textsc{t}}^{(0)}\exp(i{\bf qr}), \\
    {\bf u}_\textsc{l}^{(0)} \parallel {\bf q}, \quad {\bf u}_\textsc{t}^{(0)} \perp {\bf q}.
\end{gather}
However, this definition contains the phonon wavevector ${\bf q}$, which is ill-defined for high-frequency vibrations in amorphous medium. To generalize this approach let us note that
\begin{equation}
    \mathop\mathrm{div} {\bf u}_\textsc{t} = i {\bf qu}_\textsc{t} = 0.
\end{equation}
The transverse displacement ${\bf u}_\textsc{t}({\bf r})$ has zero divergence so it conserves local volumes.

\begin{figure}[t]
    \centerline{\includegraphics[scale=0.55]{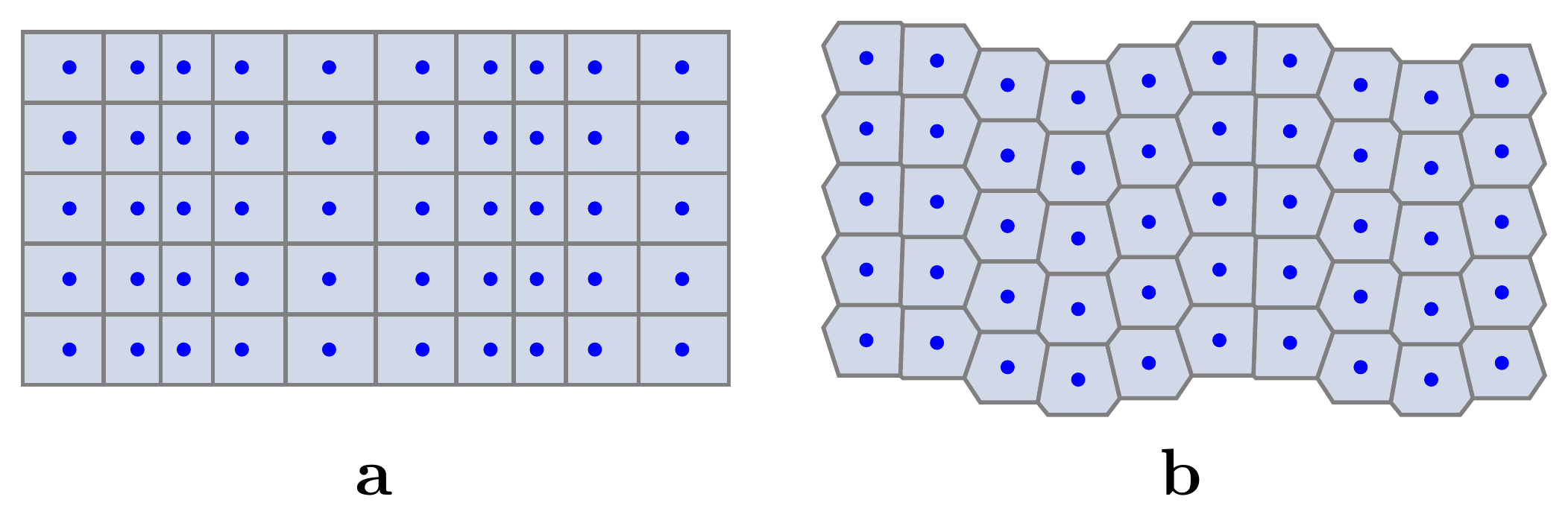}}
    \caption{Longitudinal and transverse waves in a simple quadratic lattice.}
    \label{fig:SQL}
\end{figure}

The natural analog of the local volumes in amorphous media are Voronoi cells around each atom. The Voronoi cell ${\cal V}_i$, associated with the atom $i$ is the set of all points in the space around this atom whose distance to the atom position ${\bf r}_i$ is not greater than their distance to the other atoms ${\bf r}_j$~\cite{Voronoi}. This type of cells is also known as Wigner-Seitz cells in crystallography for regular crystals. Figure~\ref{fig:SQL} shows a simple example of Voronoi cells for longitudinal and transverse waves in a simple quadratic lattice. A longitudinal wave evidently changes the volumes of Voronoi cells while the transverse wave does not change the volumes despite the change in the form of the cells.

A displacement of atoms in amorphous media may (or may not) change volumes of Voronoi cells. We will call the displacement of atoms ${\bf u}_i$ to be transverse if it does not change the volumes of all Voronoi cells. For that let us introduce a matrix $A$ which is responsible for the relative change of the $i$th Voronoi cell volume $V_i$ under $j$th atom displacement in the direction $\alpha$
\begin{equation}
    A_{i,j\alpha} = \frac{1}{V_i}\frac{\partial V_i}{\partial r_{j\alpha}}.   \label{eq:A}
\end{equation}
The explicit formula for the matrix $A$ will be derived in the next section. Using this matrix the displacement of $j$th atom in the direction $\alpha$,  $u_{j\alpha}$ produces the following relative change of the Voronoi cell volumes
\begin{equation}
    \varepsilon_i = \sum_{j\alpha}A_{i,j\alpha}u_{j\alpha}.
\end{equation}
In the matrix notation this equation reads $\varepsilon = Au$ where $A$ is a rectangular $N\times3N$ matrix (with $N$ being the number of particles) and $u$ is a displacement vector with $3N$ elements. The matrix $A$ is a discrete analog of the divergence operator.

By definition the transverse component $u_\textsc{t}$ of the displacement $u$ is such that $Au_\textsc{t}=0$, i.e. $u_\textsc{t}$ is the projection of the displacement $u$ to the \emph{null space} of the matrix $A$. The longitudinal component $u_\textsc{l}$ is a remaining orthogonal component of the displacement field and it is the projection of $u$ to the \emph{row space} of the matrix $A$. These projections have the following form~\cite{Meyer}
\begin{gather}
    u_\textsc{l} = A^T (AA^T)^{-1} A u,  \label{eq:prjns}\\
    u_\textsc{t} = u - A^T (AA^T)^{-1} A u.  \label{eq:prjrs}
\end{gather}
Indeed, $u_\textsc{t}$ does not change the Voronoi cell volumes
\begin{equation}
    A u_\textsc{t} = A u - A A^T (AA^T)^{-1} A u = 0
\end{equation}
and $u_\textsc{l}$ is orthogonal to $u_\textsc{t}$
\begin{equation}
     u_\textsc{l}^T u_\textsc{t} = u^T A^T (AA^T)^{-1} A (u - A^T (AA^T)^{-1} A u) = 0.
\end{equation}
%The formulas (\ref{eq:prjns}) and (\ref{eq:prjrs}) are universal since they do not depend on the definition of the matrix $A$ (\ref{eq:A}).

\section{The derivation of the matrix $A$}

\begin{figure}[t]
    \centerline{\includegraphics[scale=0.95]{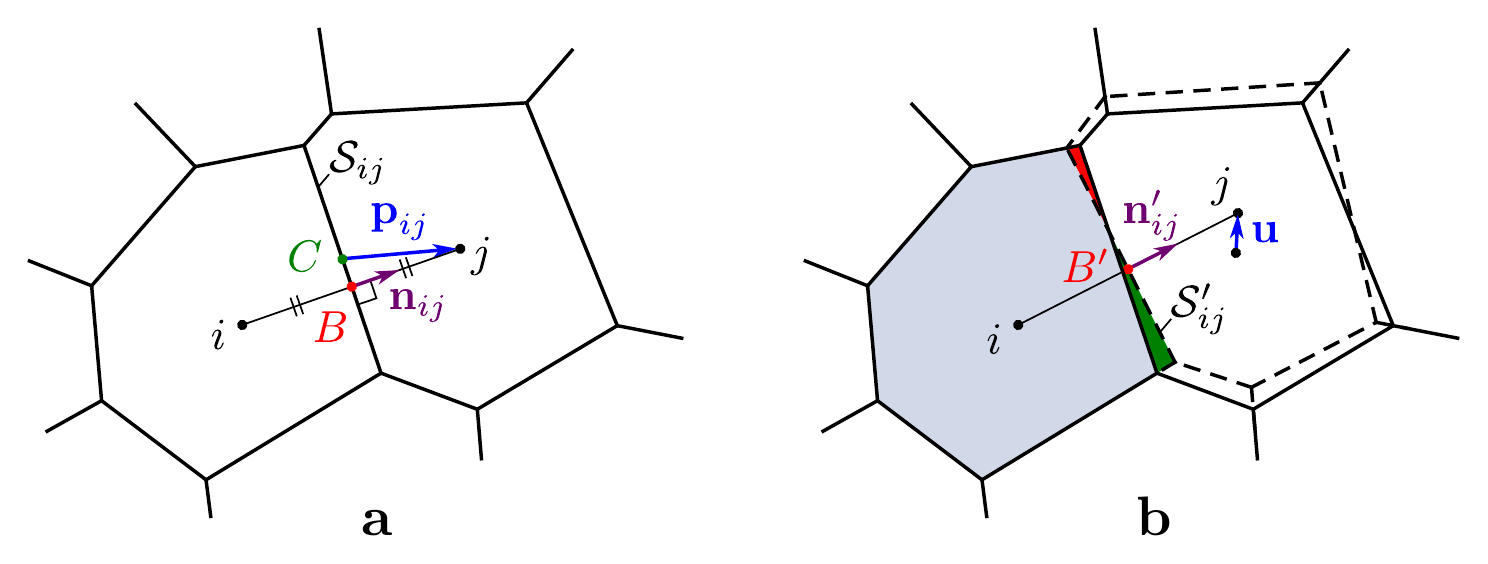}}
    \caption{a) Two dimensional example of the Voronoi cells. The points $B$ and $C$ denote the position of ${\bf b}_{ij}$ and ${\bf c}_{ij}$ respectively. In two dimensional case the point $C$ is the center of the segment ${\cal S}_{ij}$. b) Dashed lines shows the Voronoi cells after shifting of the atom $j$ by the vector ${\bf u}$. The point $B'$ denotes the position of ${\bf b}'_{ij}$. Green and red areas show the increasing and decreasing of the $V_i$ respectively.}
    \label{fig:vor}
\end{figure}

Let us show how matrix $A$ can be derived from geometry only. Let we shift only the atom $j$ by the vector ${\bf u}$ (Figure~\ref{fig:vor}). The vector ${\bf u}$ is small so the Voronoi cells after shifting have almost the same structure, but their facets are slightly shifted and rotated in space. The shifting ${\bf u}$ of the atom $j$ can change the volume $V_i$ of a nearest neighbor cell $i$. By definition the cells $i$ and $j$ are nearest neighbors if they have a common facet ${\cal S}_{ij}$.

By definition of the Voronoi cell the facet ${\cal S}_{ij}$ lies in the plane ${\cal P}_{ij}$, which has the normal ${\bf n}_{ij} = {\bf r}_{ij}/r_{ij}$, where ${\bf r}_{ij} = {\bf r}_j-{\bf r}_i$ is the vector connecting atoms $i$ and $j$ and go through the point ${\bf b}_{ij} = ({\bf r}_{i}+{\bf r}_{j})/2$. Therefore the equation for this plane reads
\begin{equation}
    {\bf n}_{ij}\cdot({\bf r}-{\bf b}_{ij}) = 0.  \label{eq:pl}
\end{equation}
After shifting of the atom $j$ the new facet ${\cal S}'_{ij}$ lies in the plane ${\cal P}'_{ij}$ which satisfy the equation
\begin{equation}
    {\bf n}'_{ij}\cdot({\bf r}-{\bf b}'_{ij}) = 0 \label{eq:pl'}
\end{equation}
where ${\bf n}'_{ij} = {\bf r}'_{ij}/r'_{ij}$, ${\bf r}'_{ij} = {\bf r}_{ij}+{\bf u}$, and ${\bf b}'_{ij} = {\bf b}_{ij} + {\bf u}/2$.
The signed distance from an arbitrary point ${\bf r}$ to the plane ${\cal P}'_{ij}$ is
\begin{equation}
    d'_{ij}({\bf r}) = {\bf n}'_{ij}\cdot({\bf b}'_{ij}-{\bf r}).
\end{equation}
This distance has the sign ``$+$'' if ${\bf r}$ lies on the same side to the plane ${\cal P}'_{ij}$ as the atom $i$ and the sign  ``$-$'' if ${\bf r}$ lies on the opposite side to the plane. If ${\bf r}$ lies in the plane ${\cal P}'_{ij}$ then the distance $d'_{ij}({\bf r})$ is equal to 0 and we get the Eq.~(\ref{eq:pl'}). In the linear approximation on $\bf u$ the change of the volume $V_i$ is the integral of $d'_{ij}({\bf r})$ over the surface of the initial facet ${\cal S}_{ij}$
\begin{gather}
    \delta V_i = \iint_{{\bf r}\in{\cal S}_{ij}} {\bf n}'_{ij}\cdot({\bf b}'_{ij}-{\bf r})\,dS = S_{ij} {\bf n}'_{ij}\cdot({\bf b}'_{ij}-{\bf c}_{ij}),  \label{eq:dVi1} \\
    {\bf c}_{ij} = \frac{1}{S_{ij}}\iint_{{\bf r}\in{\cal S}_{ij}} {\bf r}\,dS.
\end{gather}
where $S_{ij}$ and ${\bf c}_{ij}$ are correspondingly the area and the centroid of the facet ${\cal S}_{ij}$. In the linear approximation on $\bf u$, the Eq. (\ref{eq:dVi1}) reads
\begin{equation}
    \delta V_i = \frac{S_{ij}}{r_{ij}}{\bf p}_{ij}\cdot{\bf u}, \quad {\bf p}_{ij} = {\bf r}_j-{\bf c}_{ij},
\end{equation}
where we have taken into account that ${\bf r}_{ij}\cdot({\bf b}_{ij}-{\bf c}_{ij}) = 0$ because ${\bf c}_{ij}\in {\cal S}_{ij}$.

If all neighboring atoms shift, the change of the volume $V_i$ can be written in the matrix form
\begin{equation}
    \delta V_i = \sum_{j\alpha} W_{i,j\alpha}\,u_{j\alpha}.
\end{equation}
The nondiagonal elements of the matrix $W$ we have already found above
\begin{equation}
    W_{i,j\alpha} = \frac{S_{ij}}{r_{ij}}({\bf p}_{ij})_\alpha, \quad i\neq j.  \label{eq:Wij}
\end{equation}
The diagonal element  $W_{i,i\alpha}$ means the change of the $i$th Voronoi cell volume under shifting of the $i$th atom itself. Shifting of the all atoms by the same vector does not change the volumes of the Voronoi cells. Therefore
\begin{equation}
    W_{i,i\alpha} = -\sum_{j\neq i} W_{i,j\alpha}.  \label{eq:Wii}
\end{equation}
After dividing by the volume we finally get the matrix $A$
\begin{equation}
    A_{i,j\alpha} = \frac{1}{V_i}W_{i,j\alpha}.  \label{eq:Aij}
\end{equation}
It is notable that finite-elements methods also often use Voronoi cells and have a similar definition for the finite differences for the divergence operator~\cite{Mishev}.

\section{Longitudinal and transverse DOS}

\begin{figure}[t]
    \centerline{\includegraphics[scale=1.0]{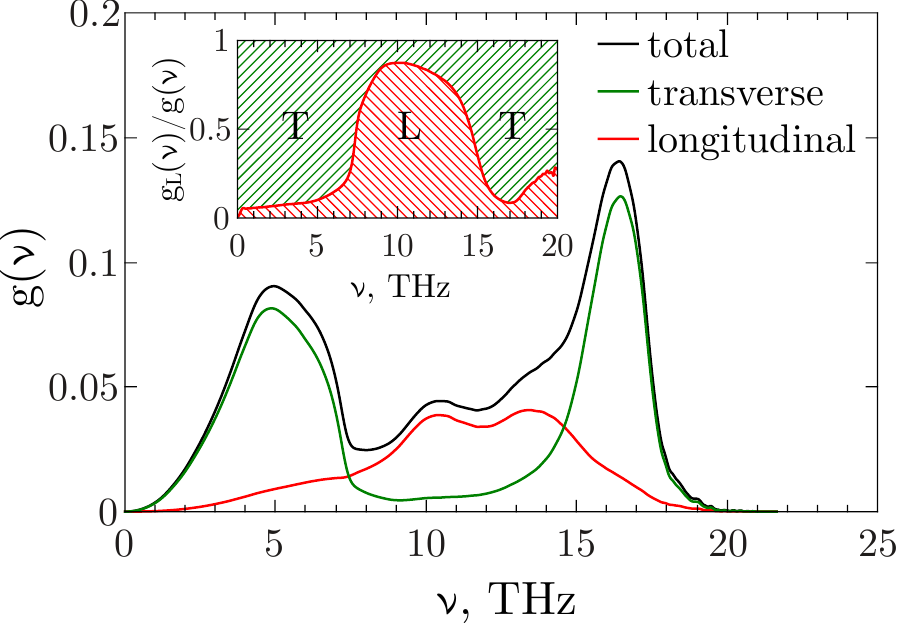}}
    \caption{The decomposition of the total vibrational density of states to longitudinal and transverse components. The inset shows the relative number of the longitudinal modes $g_\textsc{l}(\omega)/g(\omega)$ (red line). The relative number of the transverse modes $g_\textsc{t}(\omega)/g(\omega) = 1 - g_\textsc{l}(\omega)/g(\omega)$ is shown by green hatching between red line and the value 1.}
    \label{fig:DOSLT}
\end{figure}

The total vibrational density of states can be decomposed to longitudinal and transverse components
\begin{gather}
    g(\omega) = \frac{1}{3N}\sum_j \delta(\omega-\omega_j) = g_\textsc{l}(\omega) + g_\textsc{t}(\omega), \\
    g_\textsc{l,t}(\omega) = \frac{1}{3N}\sum_j \frac{\lVert u_\textsc{l,t}(\omega_j)\rVert^2}{\lVert u(\omega_j)\rVert^2}\delta(\omega-\omega_j)
\end{gather}
where $\omega_j$ is an eigenfrequency, $u(\omega_j)$ is the corresponding eigenmode, $u_\textsc{l}(\omega_j)$ and $u_\textsc{t}(\omega_j)$ are its transverse and longitudinal components defined by Eqs. (\ref{eq:prjns}) and (\ref{eq:prjrs}). The total vibrational density of states and its longitudinal and transverse components was calculated for the numerical model of a-Si consisting of $N=32768$ atoms by the Kernel polynomial method (KPM)~\cite{Weisse}. The results are shown in the Figure~\ref{fig:DOSLT}. There are three clearly seen regions. The vibrations are mostly transverse below 7 THz and above 15 THz. In the frequency interval in between the vibrations have the longitudinal nature.

The low number of longitudinal vibrations in the low-frequency region is explained by the Debye law, which gives
\begin{equation}
    g_\textsc{l}(\omega) \sim \frac{\omega^2}{c_\textsc{l}^3}, \quad g_\textsc{t}(\omega) \sim \frac{2\omega^2}{c_\textsc{t}^3}.
\end{equation}
In our model the amorphous Si has the longitudinal sound velocity $c_\textsc{l} = 7.96\ \text{km}/\text{s}$ and the transverse sound velocity $c_\textsc{t} = 3.85\ \text{km}/\text{s}$
which gives the ratio $g_\textsc{l}(\omega)/g_\textsc{t}(\omega) = c_\textsc{t}^3/2c_\textsc{l}^3 = 0.057$ for $\omega \to 0$. This value coincides with the inset in the Figure~\ref{fig:DOSLT}.

The domination of the longitudinal modes between 7 THz and 15 THz in amorphous silicon corresponds to the gap between the upper frequency of TA modes (7.5 THz) and the lower frequency of TO modes (13.9 THz) in crystalline Si~\cite{Tubino}. This frequency region in crystalline Si is totally occupied by LA and LO modes without a gap. In the same frequency region the vibrations of amorphous Si have a small transverse component (15--20\%). Certainly, there are no optical phonons with well defined wavevector in amorphous silicon due to relatively strong disorder. However, the short-range order of vibrational modes in amorphous phase can be similar to that in crystal phase. At the same time, our definition of the longitudinal and transverse vibrations is local because it is based on the Voronoi cells, which depends on the neighbor atoms only.

The Figure~\ref{fig:DOSLT} shows a sharp change of the nature of vibrations at 7 THz from almost transverse to almost longitudinal. The longitudinal vibrations correspond to the stretching of the chemical bonds between atoms while the transverse vibrations correspond to the less rigid bond bending and rocking. Therefore longitudinal vibrations can transfer the vibrational energy much faster than transverse vibrations which leads to the sudden rise of the diffusivity at 7 THz observed in Refs.~\cite{Allen-1999,Feldman-1999}. Furthermore, our additional calculations show that the longitudinal modes still have a well-defined wavevector and a sound velocity up to $\omega_\textsc{ir}^\textsc{l}\approx13$ THz unlike the transverse modes, which have the frequency of the Ioffe-Regel crossover $\omega_\textsc{ir}^\textsc{t}\approx4$ THz.

\section{Conclusion}

We have shown that vibrations in amorphous solids can be decomposed to transverse and longitudinal components in a generalized sense. The transverse component does not change the volumes of Voronoi cells around atoms while the longitudinal component is the remaining orthogonal component changing this volume. We show that the rise of the diffusivity at 7 THz in amorphous silicon corresponds to the sharp change of the nature of vibrations from almost transverse to almost longitudinal ones having high sound velocity. This decomposition can be fruitful for the investigation of the correlation function and the diffusivity of transverse and longitudinal components of vibrations in amorphous solids independently from each other.

The 3D-structure of several Voronoi cells in amorphous silicon is available online in supplementary materials.

One of the authors (YMB) thanks the Dynasty Foundation for the financial support. Two authors (DAP and YMB) thanks the University Lyon 1 for hospitality.

\end{document}